# Multiple Access Channels with States Causally Known at Transmitters

Min Li, *Member, IEEE,* Osvaldo Simeone, *Member, IEEE,* and Aylin Yener, *Member, IEEE*

*Abstract*—It has been recently shown by Lapidoth and Steinberg that strictly causal state information can be beneficial in multiple access channels (MACs). Specifically, it was proved that the capacity region of a two-user MAC with independent states, each known strictly causally to one encoder, can be enlarged by letting the encoders send compressed past state information to the decoder. In this work, a generalization of the said strategy is proposed whereby the encoders compress also the past transmitted codewords along with the past state sequences. The proposed scheme uses a combination of long-message encoding, compression of the past state sequences and codewords without binning, and joint decoding over all transmission blocks. The proposed strategy has been recently shown by Lapidoth and Steinberg to strictly improve upon the original one. Capacity results are then derived for a class of channels that include two-user modulo-additive state-dependent MACs. Moreover, the proposed scheme is extended to state-dependent MACs with an arbitrary number of users. Finally, output feedback is introduced and an example is provided to illustrate the interplay between feedback and availability of strictly causal state information in enlarging the capacity region.

## I. INTRODUCTION

State-dependent channels model relevant phenomena in wireless communication links, such as fading and interference. The standard model prescribes the existence of a state sequence $s^n = (s_1, s_2, \ldots, s_n)$, with $s_i$ denoting the state value affecting the channel at time instant $i$, with $i = 1, 2, \ldots, n$. Understanding the merit of state information, i.e., information about the sequence $s^n$, for reliable communication is a key problem of both theoretical and practical interest. In the existing literature, state-dependent channels are mainly classified into the following two groups with respect to the availability of state information at the encoders: (*i*) *non-causal* state information, where the encoders know the entire state sequence $s^n$ before encoding for the current block; (*ii*) *causal* state information, where at channel use $i$, the encoders know all states up to and including at $i$.

While referring to [2] for a thorough review on state-dependent channels, here we summarize existing results on state-dependent multiple access channels (MACs), which are the focus of our work. Reference [3] derived single-letter inner and outer bounds on the capacity region for two-user MACs with causal common state information at the encoders. Reference [4] derived a genie-aided bound to assess the capacity advantage of non-causal state over causal state information for MACs with independent state sequences available at the two encoders. Reference [5] characterized the capacity region of a cooperative MAC with state non-causally available at one encoder, while reference [6] proposed several inner and outer bounds for a MAC with states non-causally known to some encoders. A lattice coding strategy was proposed for a MAC with non-causal state information in [7] and [8].

The works summarized above demonstrate the advantages of causal and non-causal state information at the encoders for MACs. Instead, in references [9] and [10] Lapidoth and Steinberg discovered that, even with *strictly causal* state information at the encoders, an improvement in the capacity region is possible. By strictly causal state information, it is meant that at channel use $i$, the encoders know a state sequence up to, but *excluding* channel use $i$. This result stands in contrast to the well-known fact that strictly causal state information cannot improve the capacity of point-to-point channels with an independent and identically distributed (i.i.d.) state sequence. More specifically, in [9], a common state sequence is assumed to be known either strictly causally or causally at both encoders of a two-user MAC, while in [10] two independent state sequences are assumed to be available strictly causally or causally, each to a single encoder. An achievable rate region is derived in both papers and the capacity region is identified for some special cases including Gaussian models.

The main idea in the achievability proofs in [9] and [10] is to use a block Markov coding scheme in which the two users cooperatively [9] or non-cooperatively [10] transmit compressed past state information to the decoder, which in turn uses such information to perform coherent decoding. The results show that an increase in the capacity region can be obtained, even though transmission of the state information requires diverting part of the transmission resources from the transmission of message information.

Copyright (c) 2012 IEEE. Personal use of this material is permitted. However, permission to use this material for any other purposes must be obtained from the IEEE by sending a request to pubs-permissions@ieee.org.

This work was presented in part at IEEE International Symposium on Information Theory, Saint Petersburg, Russia, July 31 - August 5, 2011 [1]. The work of M. Li and A. Yener was supported in part by the National Science Foundation under Grants CNS 0716325, CCF 0964362 and DARPA ITMANET Program under Grant W911NF-07-1-0028. The work of O. Simeone was supported in part by the National Science Foundation under Grant CCF 0914899.

M. Li was with the Department of Electrical Engineering, The Pennsylvania State University, University Park, PA 16802 USA. He is now with the Department of Electronic Engineering, Macquarie University, Macquarie Park, NSW 2113, Australia (e-mail: min.li@mq.edu.au).

A. Yener is with the Department of Electrical Engineering, The Pennsylvania State University, University Park, PA 16802 USA (e-mail: yener@ee.psu.edu).

O. Simeone is with the Department of Electrical and Computer Engineering, New Jersey Institute of Technology, University Heights, NJ 07102 USA (e-mail: osvaldo.simeone@njit.edu).

## A. Contributions

In this paper, we propose a generalization of the strategy in [10] whereby the encoders compress also the past transmitted codewords along with the past state sequences. We first focus on the two-user MAC with independent states each strictly causally known to one encoder. The proposed scheme is based on long-message encoding [11], compression of the past state sequences and past codewords without binning, and joint decoding over all transmission blocks [12]. We also report on an example, put forth by Lapidoth and Steinberg in [13], in which the proposed scheme is shown to strictly improve upon the original strategy of [10].

We then generalize the capacity result for Gaussian channels of [10] for the case of a single state sequence to a larger class of channels that includes two-user modulo-additive state-dependent MACs. The proposed scheme is then extended to the state-dependent MAC with an arbitrary number of users. Finally, we introduce output feedback and show via a specific example that feedback allows user cooperation for the transmission of state information to the receiver, beside standard cooperation on the transmission of messages [14], and that this increases the capacity region.

The remainder of the paper is organized as follows: In Section II, we describe the general model considered in this work and summarize some of the existing results in [10]. Sections III and IV focus on the two-user state-dependent MAC. Section V provides a generalization to arbitrary number of users with independent states. Section VI discusses the model with output feedback. Section VII concludes the paper.

*Notation:* Throughout the paper, probability distributions are denoted by $P$ with the subscript indicating the random variables involved, e.g., $P_X(x)$ is the probability of $X = x$, and $P_{Y|X}(y|x)$ is the conditional probability of $Y = y$ given $X = x$. When the meaning is clear from the context, for convenience, we will use $P(x)$ or $P_X$ to represent $P_X(x)$. Also $x_k^i$ denotes vector $[x_{k,1}, \ldots, x_{k,i}]$. $\mathbb{E}[X]$ denotes the expectation of random variable $X$. $\mathbb{R}_i^+$ denotes the set of non-negative real vectors in $i$ dimensions. For an integer $L$, the notation $[1 : L]$ denotes the set of integers $\{1, \ldots, L\}$; for a positive real number $l$, the notation $[1 : 2^l]$ denotes the set of integers $\{1, \ldots, \lceil 2^l \rceil\}$, where $\lceil . \rceil$ is the ceiling function. In addition, $\mathcal{N}(0, \sigma^2)$ denotes a zero-mean Gaussian distribution with variance $\sigma^2$. Function $C(x)$ is defined as $C(x) = \frac{1}{2}\log_2(1+x)$.

## II. SYSTEM MODEL AND PRELIMINARIES

In this section, we describe our channel model, formulate the problem and revisit some related results derived in previous work [10].

### A. System Model

We investigate an $M$-user discrete memoryless MAC channel with $M$ mutually independent states, which is depicted in Fig. 1 and denoted by the tuple

$$\left( \begin{array}{c} \mathcal{X}_1 \times \ldots \times \mathcal{X}_M, \mathcal{S}_1 \times \ldots \times \mathcal{S}_M, \mathcal{Y}, \\ P(s_1)\ldots P(s_M), P(y|x_1, \ldots, x_M, s_1, \ldots, s_M) \end{array} \right) \quad (1)$$

with input alphabets $(\mathcal{X}_1, \ldots, \mathcal{X}_M)$, output alphabet $\mathcal{Y}$ and state alphabets $(\mathcal{S}_1, \ldots, \mathcal{S}_M)$. The state sequences are assumed to be i.i.d. and are mutually independent, i.e., $\prod_{k=1}^{M} P(s_k^n) = \prod_{k=1}^{M} \prod_{i=1}^{n} P(s_{k,i})$. The state-dependent channel is memoryless in the sense that at any discrete time $i = 1, \ldots, n$, we can write:

$$P(y_i | x_1^i, \ldots, x_M^i, s_1^i, \ldots, s_M^i, y^{i-1})$$
$$= P(y_i | x_{1,i}, \ldots, x_{M,i}, s_{1,i}, \ldots, s_{M,i}) \quad (2)$$

Each state realization is available to its corresponding encoder in a *strictly causal* manner as defined in Section I. Transmitter $k$'s signal $x_k^n$ is subject to an average input cost constraint:

$$\frac{1}{n}\sum_{i=1}^{n}\mathbb{E}[c_k(X_{k,i})] \leq \Gamma_k, \quad k = 1, \ldots, M, \quad (3)$$

where $c_k : \mathcal{X}_k \to \mathbb{R}^+$ is a single-letter input cost function for transmitter $k$ and the expectation is taken with respect to all the messages and states. We now define the following code.

*Definition 1:* Let $w_k$, uniformly distributed over the set $\mathcal{W}_k = [1 : 2^{nR_k}]$, be the message sent by transmitter $k$. A $(2^{nR_1}, \ldots, 2^{nR_M}, n, \Gamma_1, \ldots, \Gamma_M)$ code for the MAC with strictly causal and independent state information at the encoders consists of sequences of encoder mappings:

$$f_{k,i} : \mathcal{W}_k \times \mathcal{S}_k^{i-1} \to \mathcal{X}_k, \; i=1,\ldots,n, \; k=1,\ldots,M, \quad (4)$$

each of which maps message $w_k$ to a channel input such that the cost constraint (3) is satisfied, and a decoder mapping

$$g : \mathcal{Y}^n \to \mathcal{W}_1 \times \ldots \times \mathcal{W}_M, \quad (5)$$

which produces the estimate of messages $(w_1, \ldots, w_M)$.

The average probability of error, $\Pr(E)$, is defined by:

$$\Pr(E) = \prod_{k=1}^{M} 2^{-nR_k} \sum_{w_1=1}^{2^{nR_1}} \ldots \sum_{w_M=1}^{2^{nR_M}} \Pr\left( \begin{array}{c} g(y^n) \neq (w_1,\ldots,w_M) \\ |(w_1,\ldots,w_M) \; sent \end{array} \right). \quad (6)$$

Given a cost tuple $\mathbf{\Gamma} = (\Gamma_1, \ldots, \Gamma_M)$, a rate tuple $(R_1, \ldots, R_M)$ is said to be $\mathbf{\Gamma}$-achievable if there exists a sequence of codes $(2^{nR_1}, \ldots, 2^{nR_M}, n, \Gamma_1, \ldots, \Gamma_M)$ as defined above such that the probability of error satisfies $\Pr(E) \to 0$ as $n \to \infty$. The capacity region $\mathcal{C}(\mathbf{\Gamma})$ is the closure of all the $\mathbf{\Gamma}$-achievable rate tuples.

We first restrict our attention to a two-user MAC with two independent states, and then generalize to an arbitrary $M$-user MAC with $M$ independent states in Section V.

### B. Preliminaries

For comparison, we summarize a key result of [10].

*Theorem 1 ([10]):* Let $\mathbf{\Gamma} = (\Gamma_1, \Gamma_2)$ be given. Let $\mathcal{P}_{sc}$ be the set of all random variables $(V_1, V_2, S_1, S_2, X_1, X_2, Y)$ whose joint distribution is factorized as

$$P_{V_1|S_1} P_{V_2|S_2} P_{S_1} P_{S_2} P_{X_1} P_{X_2} P_{Y|S_1,S_2,X_1,X_2}. \quad (7)$$

For the two-user MAC with strictly causal state information, a $\mathbf{\Gamma}$-achievable rate region, denoted as $\mathcal{R}_{in1}(\Gamma_1, \Gamma_2)$, is given by the projection in the $(R_1, R_2)$ plane of the set of rate-cost tuples $(R_1, R_2, \Gamma_1', \Gamma_2')$ belonging to the convex hull of the collection of all the tuples $(R_1, R_2, \Gamma_1', \Gamma_2')$ satisfying

$$0 \leq R_1 \leq I(X_1; Y | X_2, V_1, V_2) - I(V_1; S_1 | Y, V_2), \quad (8a)$$
$$0 \leq R_2 \leq I(X_2; Y | X_1, V_1, V_2) - I(V_2; S_2 | Y, V_1), \quad (8b)$$
$$R_1 + R_2 \leq I(X_1, X_2; Y | V_1, V_2) - I(V_1, V_2; S_1, S_2 | Y), \quad (8c)$$
$$\text{and } \mathbb{E}[c_k(X_k)] \leq \Gamma_k', \ k = 1, 2, \quad (8d)$$

for some random variables $(V_1, V_2, S_1, S_2, X_1, X_2, Y) \in \mathcal{P}_{sc}$.

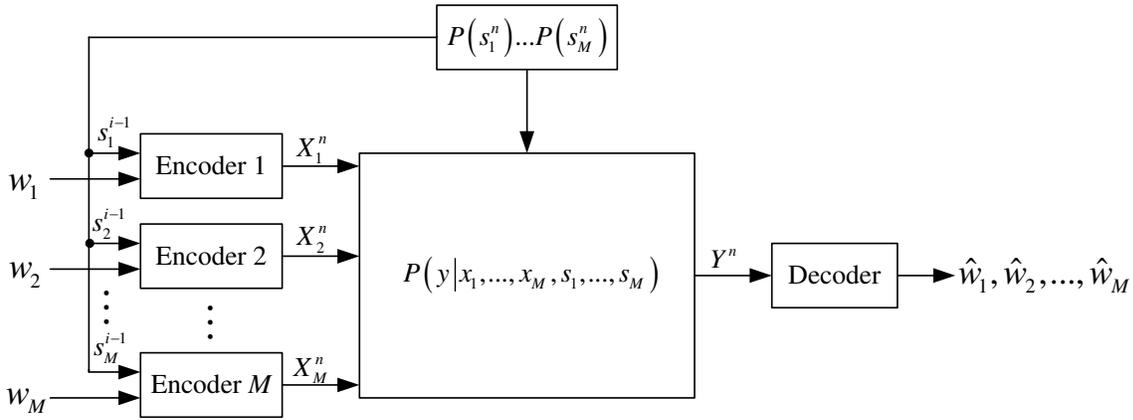

Fig. 1. The $M$-user state-dependent MAC with $M$ mutually independent states, each of which is available to its corresponding encoder in a strictly causal manner.

*Remark 1:* The basic idea of the achievable scheme of Theorem 1 is to let the transmitters convey a compressed version of the state, namely $V_1$ for $S_1$ and $V_2$ for $S_2$, to the receiver. The receiver can then use this partial information about the state to improve decoding. As an example, if the state models fading channels, state information enables partially coherent decoding. The proof of the theorem, though not available in detail in [10], is there indicated to be based on a scheme that leverages distributed Wyner-Ziv compression [15] and block Markov encoding. □

## III. A NEW ACHIEVABLE RATE REGION

In this section, for the two-user MAC ($M = 2$), we propose a new achievable scheme. The scheme is based on the idea of letting the encoders compress also the past input codewords along with the past states. We first show that the new achievable region includes the original one. Then, we report on the example put forth in [13] that demonstrates that the inclusion can be strict.

*Theorem 2:* Let $\mathbf{\Gamma} = (\Gamma_1, \Gamma_2)$ be given. Let $\mathcal{P}_{sc}^*$ be the set of all random variables $(V_1, V_2, S_1, S_2, X_1, X_2, Y)$ whose joint distribution is factorized as

$$P_{V_1|S_1,X_1} P_{V_2|S_2,X_2} P_{S_1} P_{S_2} P_{X_1} P_{X_2} P_{Y|S_1,S_2,X_1,X_2}. \quad (9)$$

For the two-user MAC with strictly causal state information, a $\mathbf{\Gamma}$-achievable rate region, denoted as $\mathcal{R}_{in2}(\Gamma_1, \Gamma_2)$, is given by the projection in $(R_1, R_2)$ plane of the set of rate-cost tuples $(R_1, R_2, \Gamma_1, \Gamma_2)$ belonging to the convex hull of the tuples $(R_1, R_2, \Gamma_1', \Gamma_2')$ satisfying

$$0 \leq R_1 < I(X_1, V_1; Y | X_2, V_2) - I(V_1; S_1 | X_1), \quad (10a)$$
$$0 \leq R_2 < I(X_2, V_2; Y | X_1, V_1) - I(V_2; S_2 | X_2), \quad (10b)$$
$$R_1 + R_2 < I(X_1, X_2, V_1, V_2; Y) - I(V_1; S_1 | X_1) - I(V_2; S_2 | X_2), \quad (10c)$$
$$\text{and } \mathbb{E}[c_k(X_k)] \leq \Gamma_k', \ k = 1, 2, \quad (10d)$$

for some random variables $(V_1, V_2, S_1, S_2, X_1, X_2, Y) \in \mathcal{P}_{sc}^*$.

*Proof:* The theorem follows as a special case of the $M$-user result of Theorem 5 for $M = 2$. We refer the reader to Appendix B for a proof of Theorem 5. ■

*Remark 2:* In the proposed strategy, the transmitters convey codewords $V_1$ and $V_2$, which compress both the past state sequences and the past transmitted codewords. This difference with respect to Theorem 1 is reflected in the different factorizations (7) and (9). Specifically, in the latter, the test channels $P_{V_k|S_k,X_k}$, $k = 1, 2$, is made to depend also on the previously transmitted symbols $X_k$. We also note that, unlike [10], our scheme uses long-message encoding, quantization without binning and joint decoding over all blocks of transmission, similar to [12] (see also [11]). □

While the joint distribution factorization (9) is more general than the original (7) used in [10], the two regions (8) and (10) are not immediately comparable given the different mutual information expressions. The next theorem shows that in fact the proposed achievable region always includes the original.

*Theorem 3:* The achievable rate region of Theorem 2 includes the achievable rate region of Theorem 1, i.e., $\mathcal{R}_{in2}(\Gamma_1, \Gamma_2) \supseteq \mathcal{R}_{in1}(\Gamma_1, \Gamma_2)$.

*Proof:* Given any cost-constraint pair $(\Gamma_1, \Gamma_2)$, setting $P_{V_1|S_1,X_1} = P_{V_1|S_1}$ and $P_{V_2|S_2,X_2} = P_{V_2|S_2}$ in $\mathcal{R}_{in2}(\Gamma_1, \Gamma_2)$, we obtain the following,

1) For the sum-rate bound,

$$R_1 + R_2$$
$$< I(X_1, X_2; Y | V_1, V_2) + I(V_1, V_2; Y)$$
$$\quad - I(V_1; S_1 | X_1) - I(V_2; S_2 | X_2) \quad (11a)$$
$$= I(X_1, X_2; Y | V_1, V_2) + H(V_1 | S_1)$$

$$+ H(V_2 | S_2) - H(V_1, V_2 | Y) \quad \text{(11b)}$$
$$= I(X_1, X_2; Y | V_1, V_2) + H(V_1 | S_1, S_2, Y)$$
$$+ H(V_2 | S_2, S_1, V_1, Y) - H(V_1, V_2 | Y) \quad \text{(11c)}$$
$$= I(X_1, X_2; Y | V_1, V_2) - I(V_1, V_2; S_1, S_2 | Y), \quad \text{(11d)}$$

where (11c) follows from the Markov chain $(V_1, V_2) \leftrightarrow (S_1, S_2) \leftrightarrow Y$ and from the fact that $(V_1, S_1)$ are independent of $(V_2, S_2)$. Note the last equation is exactly the same sum-rate bound in $\mathcal{R}_{in1}(\Gamma_1, \Gamma_2)$ given by (8c).

2) For the individual rate bound on $R_1$, we can write

$$R_1 < I(X_1; Y | X_2, V_1, V_2)$$
$$+ I(V_1; Y | X_2, V_2) - I(V_1; S_1 | X_1) \quad \text{(12a)}$$
$$= I(X_1; Y | X_2, V_1, V_2)$$
$$+ H(V_1 | S_1) - H(V_1 | Y, X_2, V_2) \quad \text{(12b)}$$
$$\geq I(X_1; Y | X_2, V_1, V_2)$$
$$+ H(V_1 | S_1) - H(V_1 | Y, V_2) \quad \text{(12c)}$$
$$= I(X_1; Y | X_2, V_1, V_2)$$
$$+ H(V_1 | S_1, Y, V_2) - H(V_1 | Y, V_2) \quad \text{(12d)}$$
$$= I(X_1; Y | X_2, V_1, V_2) - I(V_1; S_1 | Y, V_2), \quad \text{(12e)}$$

where (12c) follows from conditioning reduces entropy while (12d) follows from the Markov chain $V_1 \leftrightarrow S_1 \leftrightarrow Y$. The last equation is exactly the same as the bound on $R_1$ in $\mathcal{R}_{in1}(\Gamma_1, \Gamma_2)$ given by (8a).

3) A similar observation holds for $R_2$ by symmetry.

These three facts imply the relationship $\mathcal{R}_{in2}(\Gamma_1, \Gamma_2) \supseteq \mathcal{R}_{in1}(\Gamma_1, \Gamma_2)$. ∎

It was recently shown in [13] that the proposed region $\mathcal{R}_{in2}(\Gamma_1, \Gamma_2)$ strictly includes the original region $\mathcal{R}_{in1}(\Gamma_1, \Gamma_2)$ for some channels. The following is the example given in [13] that illustrates such inclusion.

*Example 1 ([13]):* Consider a MAC with two binary inputs $\mathcal{X}_1 = \mathcal{X}_2 = \{0, 1\}$; state $S_1 = \emptyset$ and state $S_2 = (T_0, T_1) \in \{0, 1\}^2$, where $T_0$ and $T_1$ are independent with entropies

$$H(T_0) = H(T_1) = \frac{1}{2}; \quad \text{(13)}$$

and the output $Y = (Y_1, Y_2) \in \{0, 1\}^2$ is given as

$$Y_1 = X_1 \oplus T_{X_2}, \quad \text{(14a)}$$
$$Y_2 = X_2, \quad \text{(14b)}$$

where notation "$\oplus$" denotes the conventional modulo-sum operation. The key point of this example is that the state sequence affects the received signal in a way that depends on the transmitted symbol $X_2$. Therefore, joint compression both the past state and the past codeword, or compression of the past state in way that depends on the past codeword, is expected to be beneficial. To show this, following [13], it can be seen that rate pair $(1, \frac{1}{2})$ lies in the inner bounds of $\mathcal{R}_{in2}$ by setting $V_1 = \emptyset$, $V_2 = T_{X_2}$ in (10). However, with $R_1 = 1$, it was demonstrated in [13] that $R_2$ is necessarily *zero* in $\mathcal{R}_{in1}$. This allows us to conclude, along with Theorem 3, that the region $\mathcal{R}_{in2}$ is strictly larger than the region $\mathcal{R}_{in1}$ for this example.

## IV. CAPACITY RESULT

In this section, we generalize the capacity result derived in [10] for Gaussian channels with a single state sequence to a larger class of channels.

Consider a class of discrete memoryless two-user deterministic MACs denoted by $\mathcal{D}_{MAC}$, in which the output $Y$ is a deterministic function of inputs $X_1$, $X_2$ and the channel state $S$ as

$$Y = f(X_1, X_2, S), \quad \text{(15)}$$

and where the channel state $S$, strictly causally known to encoder 1, can be calculated as a deterministic function of the inputs $X_1$, $X_2$ and the output $Y$ as

$$S = g(X_1, X_2, Y). \quad \text{(16)}$$

Then the capacity region for the class of channels $\mathcal{D}_{MAC}$ is identified as follows.

*Theorem 4:* Let $\mathbf{\Gamma} = (\Gamma_1, \Gamma_2)$ be given. For any MAC in the class $\mathcal{D}_{MAC}$ defined above, the capacity region $\mathcal{C}(\mathbf{\Gamma})$ is given by:

$$\mathcal{C}(\mathbf{\Gamma}) \triangleq \bigcup \left\{ \begin{array}{l} (R_1, R_2) \in \mathbb{R}_2^+ : \\ R_1 \leq H(Y | X_2, Q) - H(S) \\ R_2 \leq H(Y | X_1, S, Q) \\ R_1 + R_2 \leq H(Y | Q) - H(S) \end{array} \right\} \quad \text{(17)}$$

where the union is taken over all product input distributions $P_{X_1|Q} P_{X_2|Q} P_Q$ satisfying $\mathbb{E}[c_k(X_k)] \leq \Gamma_k$, $k = 1, 2$, and $Q$ is an auxiliary random variable with cardinality bound $|\mathcal{Q}| \leq 5$.

*Proof:* See Appendix A. ∎

*Remark 3:* The achievability proof in Appendix A is based on setting $V_1 = S_1 = S$ in the achievable region $R_{in2}$ in Theorem 2, which implies that $V_1$ is independent of $X_1$. Hence, the achievable scheme proposed in [10] is also optimal for the class of channels considered here. □

*Remark 4:* The class $\mathcal{D}_{MAC}$ includes the Gaussian model considered in [10], which is defined as $Y = X_1 + X_2 + S$ with input power constraints $\frac{1}{n}\sum_{i=1}^n \mathbb{E}[X_{k,i}^2] \leq P_k$ and state $S \sim \mathcal{N}(0, \sigma_s^2)$ known strictly causally to encoder 1. The capacity region $\mathcal{C}^{snf}$ for this model is given by:

$$\mathcal{C}^{snf} = \left\{ \begin{array}{l} (R_1, R_2) \in \mathbb{R}_2^+ : \\ R_1 \leq C(\frac{P_1}{\sigma_s^2}) \\ R_1 + R_2 \leq C(\frac{P_1 + P_2}{\sigma_s^2}) \end{array} \right\}. \quad \text{(18)}$$

This region can be identified from Theorem 4 by the standard extension to continuous alphabets (see, e.g., [16, Chapter 3]) and by maximizing each bound via the maximum entropy theorem [17]. Note that when providing both $S$ and $X_1$ to the receiver, the channel from user 2 to the receiver is noiseless and hence the individual bound on $R_2$ is redundant. □

*Remark 5:* The class $\mathcal{D}_{MAC}$ contains more channels along with the Gaussian model discussed in Remark 4. In particular, consider a class of binary modulo-additive state-dependent MAC channels, e.g., $Y = X_1 \oplus X_2 \oplus S$, where $S \sim \text{Bernoulli}(p_s)$, with input cost constraints $\frac{1}{n}\sum_{i=1}^n \mathbb{E}[X_{1,i}] \leq p_1$ and $\frac{1}{n}\sum_{i=1}^n \mathbb{E}[X_{2,i}] \leq p_2$, $0 < p_1, p_2, p_s \leq \frac{1}{2}$. Note that assumption (16) automatically holds for this class of binary

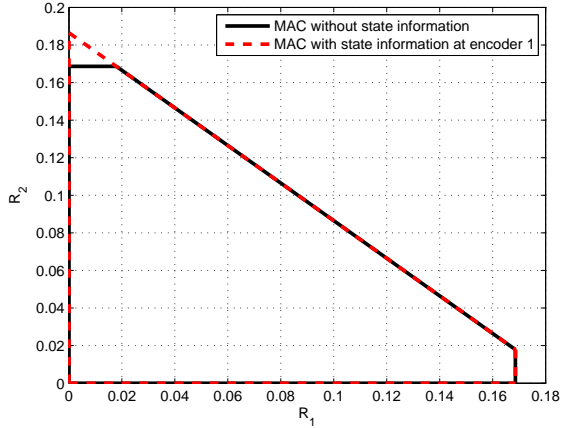

Fig. 2. Capacity region for the binary modulo-additive state-dependent MAC with input constraints considered in Remark 5 ($p_1 = p_2 = 1/3, p_s = 1/4$).

deterministic channels. From Theorem 4, by direct evaluation, we obtain that the capacity region is:

$$\mathcal{C}_{bin}^s = \left\{ \begin{array}{l} (R_1, R_2) \in \mathbb{R}_2^+ : \\ R_1 \leq H_b(p_1 * p_s) - H_b(p_s) \\ R_2 \leq H_b(p_2) \\ R_1 + R_2 \leq H_b(p_1 * p_2 * p_s) - H_b(p_s) \end{array} \right\} \quad (19)$$

where $p_1 * p_2$ denotes the convolution operation of two Bernoulli distributions with parameters $p_1$ and $p_2$, i.e., $p_1 * p_2 = p_1(1-p_2) + p_2(1-p_1)$, and $H_b(p) = -p \log_2 p - (1-p) \log_2(1-p)$. It is known from [17] that, without state information, the capacity region for this MAC channel is given by:

$$\mathcal{C}_{bin}^{ns} = \left\{ \begin{array}{l} (R_1, R_2) \in \mathbb{R}_2^+ : \\ R_1 \leq H_b(p_1 * p_s) - H_b(p_s) \\ R_2 \leq H_b(p_2 * p_s) - H_b(p_s) \\ R_1 + R_2 \leq H_b(p_1 * p_2 * p_s) - H_b(p_s) \end{array} \right\}. \quad (20)$$

Hence, we have the relationship $\mathcal{C}_{bin}^{ns} \subseteq \mathcal{C}_{bin}^s$, which confirms the benefit of strictly causal state information in enlarging the capacity region for this channel. For a numerical example, we set $p_1 = p_2 = 1/3$ and $p_s = 1/4$. The corresponding regions (19) and (20) are depicted and compared in Fig. 2. It is seen that the presence of strictly causal state information at encoder 1 improves the maximum rate of user 2. □

## V. GENERALIZATION TO $M$ USERS WITH INDEPENDENT STATES

In this section, we generalize the proposed achievable scheme to an arbitrary number $M$ of users with independent states, as depicted in Fig. 1 and described in Section II.

Let $\mathcal{A}$ denote any subset of the set of encoders $[1:M]$, i.e., $\mathcal{A} \subseteq [1:M]$ and $\mathcal{A}^c$ be the complement of $\mathcal{A}$ with respect to the set $[1:M]$. Define $\mathbf{X}(\mathcal{A})$ to be the set of random variables $X_k$ indexed by $k \in \mathcal{A}$ and similarly for $\mathbf{V}(\mathcal{A})$.

*Theorem 5:* Let cost tuple $\mathbf{\Gamma} = (\Gamma_1, \ldots, \Gamma_M)$ be given. Let $\mathcal{P}_{sc}^*$ be the set of all random variables $(V_1, \ldots, V_M, S_1, \ldots, S_M, X_1, \ldots, X_M, Y)$ whose joint distribution is factorized as

$$\prod_{k=1}^M \left( P_{V_k | S_k, X_k} P_{S_k} P_{X_k} \right) P_{Y | S_1, \ldots, S_M, X_1, \ldots, X_M}. \quad (21)$$

For the $M$-user MAC with strictly causal and independent state information, a $\mathbf{\Gamma}$-achievable rate region, denoted as $\mathcal{R}_{in}^M(\Gamma_1, \ldots, \Gamma_M)$, is given by the projection in the space $(R_1, \ldots, R_M)$ of the set of rate-cost tuples $(R_1, \ldots, R_M, \Gamma_1, \ldots, \Gamma_M)$ belonging to the convex hull of the tuples $(R_1, \ldots, R_M, \Gamma_1', \ldots, \Gamma_M')$ satisfying

$$0 \leq \sum_{k \in \mathcal{T}} R_k < \min_{\substack{\mathcal{S} \subseteq [1:M]: \\ \mathcal{T} \subseteq \mathcal{S}}} \left( \begin{array}{c} I(\mathbf{X}(\mathcal{S}), \mathbf{V}(\mathcal{S}); Y | \mathbf{X}(\mathcal{S}^c), \mathbf{V}(\mathcal{S}^c)) \\ - \sum_{l \in \mathcal{S}} I(V_l; S_l | X_l) \end{array} \right), \quad (22a)$$

$$\forall \mathcal{T} \subseteq [1:M], \quad (22b)$$

and $\mathbb{E}[c_k(X_k)] \leq \Gamma_k'$, $k = 1, \ldots, M$, \quad (22c)

for some random variables

$$(V_1, \ldots, V_M, S_1, \ldots, S_M, X_1, \ldots, X_M, Y) \in \mathcal{P}_{sc}^*.$$

*Proof:* See Appendix B. ∎

## VI. INTRODUCING OUTPUT FEEDBACK

In this section, we briefly consider an extension of the model with independent states studied in Section III, where output feedback is available to some encoder in addition to strictly causal state information. It is well known that the use of output feedback can enlarge the capacity region in MACs by allowing cooperation in the transmission of the encoders' message [14], [18], [19]. Here, instead, we demonstrate that, with strictly causal state information, a different type of cooperation is enabled by feedback that concerns the transmission of the state sequence.

To this end, we focus on the two-user state-dependent Gaussian MAC shown in Fig. 3, for which the received signal is given by:

$$Y = X_1 + X_2 + S \quad (23)$$

with power constraints $\frac{1}{n} \sum_{i=1}^n \mathbb{E}\left[X_{k,i}^2\right] \leq P_k$, for $k = 1, 2$, and state $S \sim \mathcal{N}(0, \sigma_s^2)$. We assume that the state information about $S$ is known strictly causally to the first transmitter and a perfect output feedback link is available from the receiver to the second transmitter. More specifically, we have the following encoder and decoder mappings.

*Definition 2:* Let $w_k$, uniformly distributed over the set $\mathcal{W}_k = [1 : 2^{nR_k}]$, be the message sent by transmitter $k$, $k = 1, 2$. A $(2^{nR_1}, 2^{nR_2}, n, P_1, P_2)$ code for the MAC with strictly causal state information at encoder 1 and output feedback to encoder 2 consists of the sequences of encoder mappings:

$$f_{1,i} : \mathcal{W}_1 \times \mathcal{S}^{i-1} \to \mathcal{X}_1, \quad (24a)$$

$$f_{2,i} : \mathcal{W}_2 \times \mathcal{Y}^{i-1} \to \mathcal{X}_2, \; i = 1, \ldots, n, \quad (24b)$$

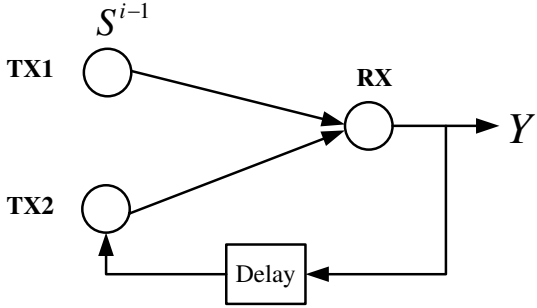

Fig. 3. The state-dependent MAC with strictly causal state information at TX1 and output feedback at TX2.

such that power constraints, i.e., $\frac{1}{n}\sum_{i=1}^n \mathbb{E}\left[X_{k,i}^2\right] \leq P_k$, for $k=1,2$, are satisfied and a decoder mapping

$$g: \mathcal{Y}^n \to \mathcal{W}_1 \times \mathcal{W}_2. \quad (25)$$

Achievability and capacity region are defined in the usual way, see Section II.

*Theorem 6:* The capacity region of the model in Fig. 3 is given by:

$$\mathcal{C}^{sf} = \bigcup_{0 \leq \rho \leq 1} \left\{ \begin{array}{l} (R_1, R_2) \in \mathbb{R}_2^+ : \\ R_1 \leq C\left(\frac{(1-\rho^2)P_1}{\sigma_s^2}\right), \\ R_1 + R_2 \leq C\left(\frac{P_1+P_2+2\rho\sqrt{P_1P_2}}{\sigma_s^2}\right) \end{array} \right\}. \quad (26)$$

*Proof:* See Appendix C. ∎

*Remark 6:* Without feedback, it is known from [9] that, if the state is known strictly causally to both encoders, the capacity is given by:

$$\mathcal{C}^{ss} = \left\{ \begin{array}{l} (R_1, R_2) \in \mathbb{R}_2^+ : \\ R_1 + R_2 \leq C\left(\frac{P_1+P_2+2\sqrt{P_1P_2}}{\sigma_s^2}\right) \end{array} \right\}, \quad (27)$$

whereas if the state is known strictly causally only to encoder 1, the capacity region $\mathcal{C}^{snf}$ is given by (18). We plot a instance of these three capacity regions by setting $P_1 = P_2 = 2$ and $\sigma_s^2 = 1$ in Fig. 4. As we observe, we have the inclusion relationships $\mathcal{C}^{snf} \subset \mathcal{C}^{sf} \subset \mathcal{C}^{ss}$. As it will be seen in the achievability proof in Appendix C, the gains obtained by leveraging feedback can be ascribed to the fact that feedback enables cooperation between the encoders in transmitting the state information to the decoder. As a further remark, consider a fourth setting in which no state information is present at encoder 1 but output feedback is available to encoder 2. While the capacity region of the case is unknown in general, it can be easily seen that $R_2 \leq C\left(\frac{P_2}{\sigma_s^2}\right)$ holds for any coding scheme. This is because the capacity of user 2 cannot be improved via feedback. Therefore, the capacity region in this case is strictly smaller than the capacity region $\mathcal{C}^{sf}$ for the case in which the state is known at encoder 1. This demonstrates the interplay between the availability of strictly causal side information at encoder 1 and of output feedback at encoder 2 in increasing the capacity region. □

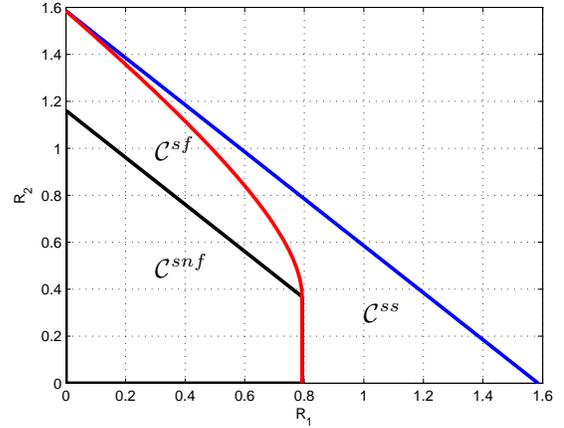

Fig. 4. Comparison of different capacity regions.

## VII. CONCLUSIONS

In this work, we have studied the state-dependent MAC with strictly causal state information at the encoders, following the original work by Lapidoth and Steinberg in [9] and [10]. We have generalized the coding scheme proposed in [10] by allowing the encoders to compress jointly past states and codewords. The proposed scheme is shown to perform at least as well as the original one, and it was demonstrated in [13] that there are channels for which it outperforms the original strategy of [10]. Moreover, the capacity result for the Gaussian model of [10] for the special case of a single state sequence has been generalized to a larger class of channels that includes two-user modulo-additive state-dependent MACs. Next, the proposed scheme has been extended to an arbitrary number of users. We have also demonstrated with an example that output feedback allows cooperation on the transmission of the state sequence in the presence of strictly causal state information. Finally, we remark that the evaluation of complete capacity region for the state-dependent MACs with strictly causal state information remains open and serves as an interesting problem for future work.

## APPENDIX A
## PROOF OF THEOREM 4

**Achievability**:

We provide the proof of achievability for $Q = q$ for a constant value $q$ and drop the conditioning on $Q$ for simplicity. The region (17) then follows by using coded time-sharing [16]. We set $V_2 = S_2 = \emptyset$ and $V_1 = S_1 = S$ in the achievable region $\mathcal{R}_{in2}$ and use the properties (15) and (16) that characterize the class of $\mathcal{D}_{MAC}$ to obtain that a rate pair $(R_1, R_2)$ is achievable if

$$R_1 < I(X_1, S; Y | X_2) - I(S; S | X_1) \quad (28a)$$
$$= H(Y | X_2) - H(S), \quad (28b)$$
$$R_2 < I(X_2; Y | X_1, S) \quad (28c)$$
$$= H(Y | X_1, S), \quad (28d)$$
$$R_1 + R_2 < I(X_1, X_2, S; Y) - I(S; S | X_1) \quad (28e)$$
$$= H(Y) - H(S) \quad (28f)$$

and $\mathbb{E}[c_k(X_k)]) \leq \Gamma_k$, $k = 1, 2$, are satisfied.

**Converse**:

From Proposition 1 and 2 in [10], we have the bounds

$$R_1 \leq I(X_1; Y | X_2, Q) + \epsilon_n \tag{29a}$$
$$= H(Y | X_2, Q) - H(S) + \epsilon_n \tag{29b}$$

and

$$R_1 + R_2 \leq I(X_1, X_2; Y | Q) + \epsilon_n \tag{30a}$$
$$= H(Y | Q) - H(S) + \epsilon_n \tag{30b}$$

where $\epsilon_n \to 0$ as $n \to \infty$, and we have defined $Q$ as a uniformly distributed random variable in the set $[1 : n]$ and independent of all other variables, and also the variables $X_1 = X_{1Q}$, $X_2 = X_{2Q}$, $Y = Y_Q$ and $S = S_Q$. Moreover, by providing perfect state information to the receiver, one can prove the following bound by using standard arguments:

$$R_2 \leq I(X_2; Y | X_1, S, Q) + \epsilon_n \tag{31a}$$
$$= H(Y | X_1, S, Q) + \epsilon_n. \tag{31b}$$

From the definition of the code, it can be seen that the distribution on $(Q, S, X_1, X_2, Y)$ is of the form $P_{Q,S,X_1,X_2,Y} = P_Q P_{X_1|Q} P_{X_2|Q} P_S P_{Y|X_1,X_2,S}$. Notice that both (29b) and (30b) leverage property (16) and the fact that $S$ is independent of $(Q, X_1, X_2)$. For the cost constraints, starting from the definition (3), we easily obtain that $\Gamma_k \geq \mathbb{E}[c_k(X_k)]$.

Finally, by the Fenchel-Eggleston-Caratheodory theorem [16, Page 631], we establish the cardinality bound $|\mathcal{Q}| \leq 5$ by observing that the rate region in Theorem 4 is characterized by the following five continuous functions over the connected compact subset given by the product probability mass functions on $\mathcal{X}_1 \times \mathcal{X}_2$: $H(Y | X_2, Q = q)$, $H(Y | X_1, S, Q = q)$, $H(Y | Q = q)$, $\mathbb{E}[c_1(X_1) | Q = q]$, and $\mathbb{E}[c_2(X_2) | Q = q]$.

## APPENDIX B
## PROOF OF THEOREM 5

Throughout the achievability proof, we use the definition of typical sequences and typical sets as in reference [16]. The set of jointly $\epsilon$-typical sequences according to a joint probability distribution $P_{X,Y}$ is denoted by $T_\epsilon^n(XY)$. When the distribution with respect to which typical sequences are defined is clear from the context, we will use $T_\epsilon^n$ for short. Throughout, we use capital letters to denote random variables and the corresponding lowercase letters to denote realized values.

In the proposed scheme, transmission takes place in $b$ blocks of $n$ channel uses each and the same message is transmitted in all blocks (long-message transmission [11]). Let $x_{k,j}^n$ be the codeword sent by user $k$ in each block $j \in [1 : b]$. This codeword encodes both user $k$'s message $w_k \in [1 : 2^{nbR_k}]$ and the index corresponding to a compressed version $v_{k,j-1}^n$ of the state sequence $s_{k,j-1}^n$ realized in the previous $(j-1)$th block and of the codeword $x_{k,j-1}^n$ transmitted in the previous block. After the $b$ transmission blocks, based on the received signals $(y_1^n, \ldots, y_b^n)$, the decoder decodes the correct message tuple $\mathbf{w} = (w_1, \ldots, w_M)$ by joint typicality decoding over all blocks. We now provide details on codebook generation, encoding and decoding operations, and probability of error analysis.

**Codebook Generation**:

Let $\epsilon > \epsilon' > 0$. Fix some probability mass function (PMF) $P_{X_k}$ such that the input cost constraint $\mathbb{E}[c_k(X_k)] \leq \Gamma_k - \epsilon$ is satisfied, and the conditional PMFs $P_{V_k|X_k,S_k}$, for all $k = 1, \ldots, M$. Define the marginal PMF $P_{V_k|X_k}(v_k|x_k) = \sum_{S_k \in \mathcal{S}_k} (P_{V_k|X_k,S_k}(v_k|x_k,s_k) P_{S_k}(s_k))$, for $k = 1, \ldots, M$. Finally, fix rate $\tilde{R}_1, \ldots, \tilde{R}_M$ (to be specified below).

1) For each block $j \in [1 : b]$, randomly and independently generate $2^{nbR_k} \times 2^{n\tilde{R}_k}$ i.i.d. sequences $x_{k,j}^n$ according to the PMF $P_{X_{k,j}^n}(x_{k,j}^n) = \prod_{i=1}^n P_{X_k}(x_{k,j,i})$, for $k = 1, \ldots, M$. Index the sequences as $x_{k,j}^n(w_k, t_{k,j-1})$, with $w_k \in [1 : 2^{nbR_k}]$ and $t_{k,j-1} \in [1 : 2^{n\tilde{R}_k}]$. As it will be discussed below, index $t_{k,j-1}$ is used to encode a compressed version of past state and transmitted codeword from a codebook of rate $\tilde{R}_k$.

2) For each block $j \in [1 : b]$ and for each codeword $x_{k,j}^n(w_k, t_{k,j-1})$, randomly and independently generate $2^{n\tilde{R}_k}$ i.i.d. sequences $v_{k,j}^n$ according to the marginal PMF $P_{V_{k,j}^n|X_{k,j}^n}(v_{k,j}^n|x_{k,j}^n) = \prod_{i=1}^n P_{V_k|X_k}(v_{k,j,i}|x_{k,j,i})$, for $k = 1, \ldots, M$. Index the sequences as $v_{k,j}^n(t_{k,j}|w_k, t_{k,j-1})$, with $t_{k,j} \in \left[1 : 2^{n\tilde{R}_k}\right]$.

**Encoding**:

Let $w_k$ be the message sent by user $k$, where $k = 1, \ldots, M$. For block $j = 1$, codeword $x_{k,1}^n(w_k, 1)$ is transmitted by user $k$. For block $j \in [2 : b]$, instead, encoder $k$ looks for an index $t_{k,j-1}$ such that

$$\begin{pmatrix} s_{k,j-1}^n, v_{k,j-1}^n(t_{k,j-1}|w_k, t_{k,j-2}), \\ x_{k,j-1}^n(w_k, t_{k,j-2}) \end{pmatrix} \in T_{\epsilon'}^n(S_k V_k X_k). \tag{32}$$

If no such index is found, then an arbitrary index $t_{k,j-1}$ is selected from the set $[1 : 2^{n\tilde{R}_k}]$. If more than one such index is found, the first one in lexicographical order is selected. Finally, the codeword $x_{k,j}^n(w_k, t_{k,j-1})$ is transmitted by user $k$ in the $j$th block.

**Decoding**:

After $b$ blocks of transmission, the decoder looks for a unique message tuple $\hat{\mathbf{w}} = (\hat{w}_1, \ldots, \hat{w}_M)$, where $\hat{w}_k \in [1 : 2^{nbR_k}]$, such that there exists *some* tuple $(t_{1,j}, \ldots, t_{M,j})$, with $t_{k,j} \in [1 : 2^{n\tilde{R}_k}]$ and $j \in [1 : b]$, satisfying the condition

$$\begin{pmatrix} x_{1,j}^n(\hat{w}_1, t_{1,j-1}), \ldots, x_{M,j}^n(\hat{w}_M, t_{M,j-1}), \\ v_{1,j}^n(t_{1,j}|\hat{w}_1, t_{1,j-1}), \ldots, v_{M,j}^n(t_{M,j}|\hat{w}_M, t_{M,j-1}), y_j^n \end{pmatrix}$$
$$\in T_\epsilon^n(X_1 \ldots X_M V_1 \ldots V_M Y) \tag{33}$$

for all blocks $j \in [1 : b]$.

**Probability of Error Analysis**:

We now bound the probability of error $\Pr(E)$ averaged over all distribution of the codebooks defined above. Without loss of generality, given the symmetry of the codebook generation, we assume the message tuple sent is $\mathbf{w} = (1, \ldots, 1) \triangleq \mathbf{1}_M$ and we label the compression index chosen by encoder $k$ for each block $j$ as $t_{k,j} = 1$. In the following, we first define the error events associated with the encoding and decoding operations, and then bound the corresponding probabilities of error.

Let $E_0 = \bigcup_{k=1}^{M} E_{0,k}$ denote the event corresponding to encoding errors, where $E_{0,k}$ represents the error event at encoder $k$, for $k = 1,\ldots,M$. An encoding error at encoder $k$ occurs when in some block $j$ there is no codeword $V_{k,j-1}^n(t_{k,j-1}|1,1)$ satisfying the joint typicality rule (32). Therefore, the error event $E_{0,k}$ can be written as the union

$$E_{0,k} = \bigcup_{j=1}^{b} \left\{ \begin{pmatrix} S_{k,j-1}^n, V_{k,j-1}^n(t_{k,j-1}|1,1), \\ X_{k,j-1}^n(1,1) \end{pmatrix} \notin T_{\epsilon'}^n, \\ \text{for all } t_{k,j-1} \in [1:2^{n\tilde{R}_k}] \right\}. \quad (34)$$

In order to define the decoding error events, we first define the event $E_\mathbf{w}$ indexed by a message tuple $\mathbf{w} = (w_1,\ldots,w_M)$ as given by (35), where we have defined that $\mathbf{t}_j = (t_{1,j}, t_{2,j},\ldots,t_{M,j})$ and $t_{k,j} \in [1:2^{n\tilde{R}_k}]$, $k = 1,\ldots,M$. Event $E_\mathbf{w}$ occurs when the decoder finds a message tuple $\mathbf{w}$ satisfying the decoding rule (33). Based on the decoding rule (33), the decoding error event can thus be expressed as the union $E_{\mathbf{1}_M}^c \bigcup \{\bigcup_{\mathbf{w}\neq \mathbf{1}_M} E_\mathbf{w}\}$.

Overall, by considering both encoding and decoding errors and leveraging the union bound, the probability of error can be upper bounded as

$$\Pr(E) \leq \sum_{k=1}^{M} \Pr(E_{0,k}) + \Pr\left(E_{\mathbf{1}_M}^c \cap E_0^c\right) + \sum_{\mathbf{w}\neq \mathbf{1}_M} \Pr(E_\mathbf{w}). \quad (36)$$

We now consider separately the terms in the sum (36).

1) By the covering lemma [16], we have the limit $\Pr(E_{0,k}) \to 0$ as long as the inequality

$$\tilde{R}_k > I(S_k; V_k|X_k) + \delta(\epsilon') \quad (37)$$

holds for sufficiently large $n$, where $\delta(\epsilon') \to 0$ as $\epsilon' \to 0$.

2) By the conditional joint typicality lemma [16], we have that $\Pr\left(E_{\mathbf{1}_M}^c \cap E_0^c\right) \to 0$ for sufficiently large $n$.

3) To bound each term in the third summand in (36), for convenience, for any given $\mathbf{w} \neq \mathbf{1}_M$, $\mathbf{t}_j = (t_{1,j},\ldots,t_{M,j})$ and $\mathbf{t}_{j-1} = (t_{1,j-1},\ldots,t_{M,j-1})$, we define the event $A_j(\mathbf{w},\mathbf{t}_j,\mathbf{t}_{j-1})$ as

$$A_j(\mathbf{w},\mathbf{t}_j,\mathbf{t}_{j-1}) = \\ \left\{ \begin{pmatrix} X_{1,j}^n(w_1,t_{1,j-1}),\ldots,X_{M,j}^n(w_M,t_{M,j-1}), \\ V_{1,j}^n(t_{1,j}|w_1,t_{1,j-1}),\ldots, \\ V_{M,j}^n(t_{M,j}|w_M,t_{M,j-1}), Y_j^n \end{pmatrix} \in T_\epsilon^n \right\}. \quad (38)$$

From (35), we have the following

$$\Pr(E_\mathbf{w}) = \Pr\left(\bigcup_{\mathbf{t}^b} \bigcap_{j=1}^b A_j(\mathbf{w},\mathbf{t}_j,\mathbf{t}_{j-1})\right) \quad (39a)$$

$$\leq \sum_{\mathbf{t}^b} \Pr\left(\bigcap_{j=1}^b A_j(\mathbf{w},\mathbf{t}_j,\mathbf{t}_{j-1})\right) \quad (39b)$$

$$\leq \sum_{\mathbf{t}^b} \prod_{j=2}^b \Pr(A_j(\mathbf{w},\mathbf{t}_j,\mathbf{t}_{j-1})), \quad (39c)$$

where the union and sums over $\mathbf{t}^b$ are taken over all vectors $\mathbf{t}^b$ as defined in (35); and (39c) holds due to the independence of the codebooks generated for each block, the memoryless property of the channel and the fact that $0 \leq \Pr(A_1) \leq 1$.

Next, we provide an upper bound on the probability $\Pr(A_j(\mathbf{w},\mathbf{t}_j,\mathbf{t}_{j-1}))$ for a given tuple $(\mathbf{w},\mathbf{t}_j,\mathbf{t}_{j-1})$. To facilitate the analysis, we introduce some useful notation. Specifically, for any given pair of vectors $(\mathbf{w},\mathbf{t}_{j-1})$ with $j \in [2:b]$, we define the index set $\mathcal{S}_j(\mathbf{w},\mathbf{t}_{j-1})$, where we will drop the dependence on the arguments where necessary to simplify the notation. This set contains all the indices $k$ for which at least one of the conditions $w_k \neq 1$ and $t_{k,j-1} \neq 1$ is satisfied for the pair of vectors $(\mathbf{w},\mathbf{t}_{j-1})$, i.e.,

$$\mathcal{S}_j(\mathbf{w},\mathbf{t}_{j-1}) = \{k \in [1:M] : w_k \neq 1 \text{ or } t_{k,j-1} \neq 1\}. \quad (40)$$

In addition, let $\mathcal{S}_j^c(\mathbf{w},\mathbf{t}_{j-1})$ denote the complement of $\mathcal{S}_j(\mathbf{w},\mathbf{t}_{j-1})$ with respect to the set $[1:M]$, i.e., $\mathcal{S}_j^c(\mathbf{w},\mathbf{t}_{j-1}) = \{k \in [1:M] \setminus \mathcal{S}_j(\mathbf{w},\mathbf{t}_{j-1})\}$. Furthermore, we partition the set $\mathcal{S}_j^c(\mathbf{w},\mathbf{t}_{j-1})$ into two subsets as follows:

$$\mathcal{S}_j'(\mathbf{w},\mathbf{t}_{j-1},\mathbf{t}_j) = \{k \in \mathcal{S}_j^c(\mathbf{w},\mathbf{t}_{j-1}) : t_{k,j} \neq 1\}, \quad (41a)$$
$$\mathcal{S}_j''(\mathbf{w},\mathbf{t}_{j-1},\mathbf{t}_j) = \{k \in \mathcal{S}_j^c(\mathbf{w},\mathbf{t}_{j-1}) : t_{k,j} = 1\}. \quad (41b)$$

By definition, we have that

$$\mathcal{S}_j'(\mathbf{w},\mathbf{t}_{j-1},\mathbf{t}_j) \bigcup \mathcal{S}_j''(\mathbf{w},\mathbf{t}_{j-1},\mathbf{t}_j) = \mathcal{S}_j^c(\mathbf{w},\mathbf{t}_{j-1}).$$

Finally, for a generic set $\mathcal{A}_j \subseteq [1:M]$, we define as $\mathbf{X}(\mathcal{A}_j)$ to be the set of variables $X_{k,j}$, for $k \in \mathcal{A}_j$, where $X_{k,j}$ is the symbol transmitted by the $k$th user in the $j$th block. We use similar definition for $\mathbf{V}(\mathcal{A}_j)$.

Given the above notation and by the codebook construction, we use standard arguments on joint typicality [16] to obtain

$$\Pr(A_j(\mathbf{w},\mathbf{t}_j,\mathbf{t}_{j-1}))$$
$$\leq 2^{-n\begin{pmatrix} H(\mathbf{X}(\mathcal{S}_j),\mathbf{V}(\mathcal{S}_j))+H(\mathbf{V}(\mathcal{S}_j')|\mathbf{X}(\mathcal{S}_j'))+H(\mathbf{X}(\mathcal{S}_j^c),\mathbf{V}(\mathcal{S}_j''),Y) \\ -H(\mathbf{X}(\mathcal{S}_j),\mathbf{X}(\mathcal{S}_j^c),\mathbf{V}(\mathcal{S}_j),\mathbf{V}(\mathcal{S}_j^c),Y)-\delta(\epsilon) \end{pmatrix}} \quad (42a)$$
$$= 2^{-n\begin{pmatrix} I(\mathbf{X}(\mathcal{S}_j),\mathbf{V}(\mathcal{S}_j);\mathbf{X}(\mathcal{S}_j^c),\mathbf{V}(\mathcal{S}_j''),Y|\mathbf{V}(\mathcal{S}_j')) \\ +I(\mathbf{V}(\mathcal{S}_j');\mathbf{X}(\mathcal{S}_j^c),\mathbf{V}(\mathcal{S}_j''),Y)-I(\mathbf{V}(\mathcal{S}_j');\mathbf{X}(\mathcal{S}_j'))-\delta(\epsilon) \end{pmatrix}} \quad (42b)$$
$$\leq 2^{-n\left(I(\mathbf{X}(\mathcal{S}_j),\mathbf{V}(\mathcal{S}_j);\mathbf{X}(\mathcal{S}_j^c),\mathbf{V}(\mathcal{S}_j''),Y|\mathbf{V}(\mathcal{S}_j'))-\delta(\epsilon)\right)} \quad (42c)$$
$$= 2^{-n\left(I(\mathbf{X}(\mathcal{S}_j),\mathbf{V}(\mathcal{S}_j);Y|\mathbf{V}(\mathcal{S}_j'),\mathbf{X}(\mathcal{S}_j^c),\mathbf{V}(\mathcal{S}_j''))-\delta(\epsilon)\right)} \quad (42d)$$
$$= 2^{-n\left(I(\mathbf{X}(\mathcal{S}_j),\mathbf{V}(\mathcal{S}_j);Y|\mathbf{X}(\mathcal{S}_j^c),\mathbf{V}(\mathcal{S}_j^c))-\delta(\epsilon)\right)}, \quad (42e)$$

$$E_\mathbf{w} = \left\{ \bigcap_{j=1}^{b} \left\{ \begin{pmatrix} X_{1,j}^n(w_1,t_{1,j-1}),\ldots,X_{M,j}^n(w_M,t_{M,j-1}), \\ V_{1,j}^n(t_{1,j}|w_1,t_{1,j-1}),\ldots,V_{M,j}^n(t_{M,j}|w_M,t_{M,j-1}), Y_j^n \end{pmatrix} \in T_\epsilon^n \right\}, \\ \text{for some } \mathbf{t}^b \triangleq (\mathbf{t}_1,\ldots,\mathbf{t}_b). \right\} \quad (35)$$

where $\delta(\epsilon) \to 0$ as $\epsilon \to 0$; (42b) follows from standard steps involving mutual information; (42c) holds because $\mathcal{S}_j' \subseteq \mathcal{S}_j^c$ so that $I(\mathbf{V}(\mathcal{S}_j'); \mathbf{X}(\mathcal{S}_j^c), \mathbf{V}(\mathcal{S}_j''), Y) \geq I(\mathbf{V}(\mathcal{S}_j'); \mathbf{X}(\mathcal{S}_j^c))$; (42d) holds because of the fact that the tuple $(\mathbf{X}(\mathcal{S}_j), \mathbf{V}(\mathcal{S}_j))$ is independent of the tuple $(\mathbf{V}(\mathcal{S}_j'), \mathbf{X}(\mathcal{S}_j^c), \mathbf{V}(\mathcal{S}_j''))$; and finally (42e) is due to the fact that $\mathcal{S}_j' \bigcup \mathcal{S}_j'' = \mathcal{S}_j^c$. It is noted that the upper bound of (42e) depends only on the sets $\mathcal{S}_j(\mathbf{w}, \mathbf{t}_{j-1})$ and $\mathcal{S}_j^c(\mathbf{w}, \mathbf{t}_{j-1})$, and hence it is independent of $\mathbf{t}_j$ for any given $\mathbf{w}$ and $\mathbf{t}_{j-1}$.

Given this upper bound, we then proceed with (39c) and obtain the following

$$\Pr(E_\mathbf{w})$$
$$\leq \sum_{\mathbf{t}^b} \prod_{j=2}^b \Pr(A_j(\mathbf{w}, \mathbf{t}_j, \mathbf{t}_{j-1})) \tag{43a}$$
$$= \sum_{\mathbf{t}_b} \sum_{\mathbf{t}^{b-1}} \prod_{j=2}^b \Pr(A_j(\mathbf{w}, \mathbf{t}_j, \mathbf{t}_{j-1})) \tag{43b}$$
$$\leq \sum_{\mathbf{t}_b} \sum_{\mathbf{t}^{b-1}} \prod_{j=2}^b 2^{-n(I(\mathbf{X}(\mathcal{S}_j), \mathbf{V}(\mathcal{S}_j); Y | \mathbf{X}(\mathcal{S}_j^c), \mathbf{V}(\mathcal{S}_j^c)) - \delta(\epsilon))} \tag{43c}$$
$$= \sum_{\mathbf{t}_b} \prod_{j=2}^b \sum_{\mathbf{t}_{j-1}} 2^{-n(I(\mathbf{X}(\mathcal{S}_j), \mathbf{V}(\mathcal{S}_j); Y | \mathbf{X}(\mathcal{S}_j^c), \mathbf{V}(\mathcal{S}_j^c)) - \delta(\epsilon))} \tag{43d}$$
$$\leq 2^{n \sum_{k \in [1:M]} \tilde{R}_k} \left( \sum_{\substack{\mathcal{S} \subseteq [1:M]: \\ \mathcal{T}(\mathbf{w}) \subseteq \mathcal{S}}} \sum_{\substack{\mathbf{t}_{j-1}: \\ \mathcal{S}_j(\mathbf{w}, \mathbf{t}_{j-1}) = \mathcal{S}}} 2^{-n(\mathbf{I}_{(\mathcal{S})} - \delta(\epsilon))} \right)^{b-1} \tag{43e}$$
$$\leq 2^{n \sum_{k \in [1:M]} \tilde{R}_k} \left( \sum_{\substack{\mathcal{S} \subseteq [1:M]: \\ \mathcal{T}(\mathbf{w}) \subseteq \mathcal{S}}} 2^{n \sum_{l \in \mathcal{S}} \tilde{R}_l} 2^{-n(\mathbf{I}_{(\mathcal{S})} - \delta(\epsilon))} \right)^{b-1} \tag{43f}$$
$$\leq 2^{n \sum_{k \in [1:M]} \tilde{R}_k} \left( 2^{(M-1)} 2^{-n(\mathbf{I}_{\min} - \delta(\epsilon))} \right)^{b-1}, \tag{43g}$$
$$= 2^{\left( n \sum_{k \in [1:M]} \tilde{R}_k + (b-1)(M-1) - n(b-1)(\mathbf{I}_{\min} - \delta(\epsilon)) \right)} \tag{43h}$$

where (43c) follows from (42e); (43d) holds because of the fact that the upper bound (42e) is independent of $\mathbf{t}_j$ for any given $\mathbf{w}$ and $\mathbf{t}_{j-1}$; (43e) also follows from (42e), where we have defined the index set $\mathcal{T}(\mathbf{w}) = \{k \in [1:M] : w_k \neq 1\}$ and $\mathbf{I}_{(\mathcal{S})} = I(\mathbf{X}(\mathcal{S}), \mathbf{V}(\mathcal{S}); Y | \mathbf{X}(\mathcal{S}^c), \mathbf{V}(\mathcal{S}^c))$; (43f) follows by $t_{l,j-1} \in [1 : 2^{n\tilde{R}_l}]$ for any $l \in \mathcal{S}$; and (43g) holds because there are at most $2^{(M-1)}$ subsets of $[1 : M]$ that contain any index set $\mathcal{T}(\mathbf{w})$ given, where we have defined the term

$$\mathbf{I}_{\min} = \min_{\substack{\mathcal{S} \subseteq [1:M]: \\ \mathcal{T}(\mathbf{w}) \subseteq \mathcal{S}}} \left( \mathbf{I}_{(\mathcal{S})} - \sum_{l \in \mathcal{S}} \tilde{R}_l \right). \tag{44}$$

In this way, we obtain that

$$\sum_{\mathbf{w} \neq \mathbf{1}_M} \Pr(E_\mathbf{w})$$
$$\leq \sum_{\mathcal{T} \subseteq [1:M]} 2^{\left( \begin{array}{c} nb \sum_{k \in \mathcal{T}} R_k + n \sum_{k \in [1:M]} \tilde{R}_k \\ + (b-1)(M-1) \\ -n(b-1)(\mathbf{I}_{\min} - \delta(\epsilon)) \end{array} \right)}. \tag{45}$$

Therefore, we conclude that the limit $\sum_{\mathbf{w} \neq \mathbf{1}_M} \Pr(E_\mathbf{w}) \to 0$ holds as long as the condition:

$$nb \sum_{k \in \mathcal{T}} R_k + n \sum_{k \in [1:M]} \tilde{R}_k + (b-1)(M-1)$$
$$< n(b-1)(\mathbf{I}_{\min} - \delta(\epsilon)), \ \forall \mathcal{T} \subseteq [1:M], \tag{46}$$

is satisfied, or equivalently we have

$$\sum_{k \in \mathcal{T}} R_k < \frac{(b-1)}{b}(\mathbf{I}_{\min} - \delta(\epsilon)) - \frac{\sum_{k \in [1:M]} \tilde{R}_k}{b}$$
$$- \frac{(b-1)(M-1)}{nb}, \ \forall \mathcal{T} \subseteq [1:M]. \tag{47}$$

Setting $b \to \infty$ and $n \to \infty$, we then have the condition

$$\sum_{k \in \mathcal{T}} R_k < \mathbf{I}_{\min}$$
$$= \min_{\substack{\mathcal{S} \subseteq [1:M]: \\ \mathcal{T} \subseteq \mathcal{S}}} \left( \begin{array}{c} I(\mathbf{X}(\mathcal{S}), \mathbf{V}(\mathcal{S}); Y | \mathbf{X}(\mathcal{S}^c), \mathbf{V}(\mathcal{S}^c)) \\ - \sum_{l \in \mathcal{S}} \tilde{R}_l \end{array} \right) \tag{48a}$$
$$\leq \min_{\substack{\mathcal{S} \subseteq [1:M]: \\ \mathcal{T} \subseteq \mathcal{S}}} \left( \begin{array}{c} I(\mathbf{X}(\mathcal{S}), \mathbf{V}(\mathcal{S}); Y | \mathbf{X}(\mathcal{S}^c), \mathbf{V}(\mathcal{S}^c)) \\ - \sum_{l \in \mathcal{S}} I(V_l; S_l | X_l) \end{array} \right) \tag{48b}$$

for all $\mathcal{T} \subseteq [1 : M]$. This completes the proof of Theorem 5.

## APPENDIX C
## PROOF OF THEOREM 6

**Achievability**:

The key idea of the achievable scheme is based on a variation of Schalkwijk-Kailath coding [14], [20]. User 1 divides its power into two parts. Specifically, it consumes fraction $\alpha$ ($0 \leq \alpha \leq 1$) of its power to send its message $w_1$ over $n$ channel uses using a codeword drawn from a codebook whose entries are generated in an i.i.d. fashion from a zero-mean Gaussian distribution with variance $\alpha P_1$. Moreover, it uses the remaining portion $(1 - \alpha) P_1$ to transmit state information via in cooperation with user 2, as detailed below.

*Codebook Generation*: Randomly generate $2^{nR_1}$ i.i.d. sequences $x_{1p}^n$ with each component distributed as $x_{1p,i} \sim \mathcal{N}(0, \alpha P_1)$, for $i = 1, \ldots, n$. Index the sequences by $x_{1p}^n(w_1)$ with $w_1 \in [1 : 2^{nR_1}]$. Partition the interval $[-1 : 1]$ into $2^{nR_2}$ small intervals of equal length and map messages $w_2 \in [1 : 2^{nR_2}]$ to the middle points of these intervals. Index these middle points by $\theta(w_2)$.

*Encoding*:
1) Initial channel use, $i = 0$: User 1 keeps silent in this channel use. To send message $w_2$ to the receiver, user 2 transmits $\theta(w_2)$;

2) First channel use, $i = 1$: By feedback, user 2 learns state $s_0$ after subtracting its own information. Since user 1 knows $s_0$ as well, it cooperates with user 2 to convey information about state $s_0$ to the receiver, superimposed on its private message $w_1$. Specifically, user 1 transmits $x_{1,1} = x_{1p,1}(w_1) + \gamma_{1,1} s_0$, where the scalar $\gamma_{1,1}$ is chosen so that $\gamma_{1,1} s_0 \sim \mathcal{N}(0, (1-\alpha)P_1)$, while user 2 transmits $x_{2,1} = \gamma_{2,1} s_0$, where the scalar $\gamma_{2,1}$ is chosen so that $\gamma_{2,1} s_0 \sim \mathcal{N}(0, P_2)$;

3) Channel uses $i \geq 2$: At each following channel use $i$, user 2 forms the minimum mean squared error (MMSE) estimate $\mathbb{E}[s_0 | y_1^{i-1}]$ of $s_0$ based on the observed output symbols $y_1^{i-1}$ at the beginning of channel use $i$. Let $s'_{i-1} = s_0 - \mathbb{E}[s_0 | y_1^{i-1}]$. Then user 2 transmits $x_{2,i} = \gamma_{2,i} s'_{i-1}$ over the channel use $i$, where the scalar $\gamma_{2,i}$ is selected so that $\gamma_{2,i} s'_{i-1} \sim \mathcal{N}(0, P_2)$. Given the fact that user 1 knows $s_0$, the outdated channel state $s^{i-1}$ and its own message symbols, it equivalently knows the channel output symbols from the first channel use up to current time. Hence it can also generate the MMSE estimate of $s_0$ and thus $s'_{i-1}$ as done by user 2. User 1 then transmits $x_{1,i} = x_{1p,i}(w_1) + \gamma_{1,i} s'_{i-1}$ in channel use $i$, where the scalar $\gamma_{1,i}$ is chosen so that $\gamma_{1,i} s'_{i-1} \sim \mathcal{N}(0, (1-\alpha)P_1)$.

*Decoding*: After $n+1$ channel uses, the receiver first estimates state $s_0$ by $\hat{s}_0 = \mathbb{E}[s_0 | y_1^n]$; it then estimates $\theta(w_2)$ by $\hat{\theta} = y_0 - \hat{s}_0 = \theta(w_2) + (s_0 - \mathbb{E}[s_0 | y_1^n])$ and declares that message $\hat{w}_2$ is sent if $\theta(\hat{w}_2)$ is the closest message point to $\hat{\theta}$. After successfully estimating state $s_0$ and decoding message $w_2$, the receiver is able to retrieve the information about $s_0$, which is conveyed from both users, so as to subtract it from the received sequence $y_1^n$. In this way, message $w_1$ is decoded based on the residual signal.

*Analysis of Probability of Error*: We note that using the union bound, we have, $\Pr(E) \leq \Pr(E_2) + \Pr(E_1 | E_2^c)$, where the first term corresponds to the probability of decoding error for message $w_2$, and the second term is the probability of decoding error for message $w_1$ given that message $w_2$ is correctly decoded. The probability of decoding error $\Pr(E_2)$ vanishes as the variance of estimation error of $s_0$ becomes arbitrarily small as $n \to \infty$. Similar to [14], it can be shown that we have $\Pr(E_2) \to 0$ as long as

$$R_2 \leq C\left(\frac{(1-\alpha)P_1 + P_2 + 2\sqrt{(1-\alpha)P_1 P_2}}{\sigma_s^2 + \alpha P_1}\right). \quad (49)$$

Moreover, from the standard consideration, we have $\Pr(E_1 | E_2^c) \to 0$ as long as the inequality

$$R_1 \leq C\left(\frac{\alpha P_1}{\sigma_s^2}\right) \quad (50)$$

holds. Setting $\alpha \triangleq (1 - \rho^2)$ concludes the proof of the achievability.

It is remarked that the achievability can also be proved by extending the scheme proposed in [9, page 15]. This scheme demonstrates that it is enough for both users to know the initial state symbol $s_0$, which can be accomplished by user 2 via feedback, in order to achieve the rate region of Theorem 6.

**Converse**:

Providing message $w_2$ to encoder 1, the channel becomes the MAC studied in [21] where encoder 1 knows both $w_1$ and $w_2$, encoder 2 knows $w_2$ and output feedback is available at the encoders. In fact, the state sequence at encoder 1 in this genie-aided model can be seen as equivalent to feedback, since via feedback, encoder 1 effectively obtains $s^{i-1}$. It is shown in [21] that feedback does not increase capacity and thus the capacity region is given by (26).

## ACKNOWLEDGMENT

The authors would like to thank the Associate Editor and the anonymous reviewers for their comments, which have improved the quality of the paper.

**Min Li** (M'12) received his B.E. degree in Telecommunications Engineering from Zhejiang University, Hangzhou, China in June 2006, and his M.E. degree in Information and Communication Engineering from Zhejiang University, Hangzhou, China in June 2008. He received his Ph.D. degree in Electrical Engineering from The Pennsylvania State University, University Park, United States in August 2012. Since September 2012, he is a research fellow in wireless communications at the Department of Electronic Engineering at Macquarie University, Australia.

His research interests include network information theory, coding theory, wireless communication theory and system designs, optimization techniques and VLSI designs.

**Osvaldo Simeone** (M'02) received the M.Sc. degree (with honors) and the Ph.D. degree in information engineering from Politecnico di Milano, Milan, Italy, in 2001 and 2005, respectively.

He is currently with the Center for Wireless Communications and Signal Processing Research (CWCSPR), New Jersey Institute of Technology (NJIT), Newark, where he is an Associate Professor. His current research interests concern the cross-layer analysis and design of wireless networks with emphasis on information-theoretic, signal processing, and queuing aspects. Specific topics of interest are: cognitive radio, cooperative communications, rate-distortion theory, ad hoc, sensor, mesh and hybrid networks, distributed estimation, and synchronization.

Dr. Simeone is a co-recipient of Best Paper Awards of the IEEE SPAWC 2007 and IEEE WRECOM 2007. He currently serves as an Editor for IEEE TRANSACTIONS ON COMMUNICATIONS.

**Aylin Yener** (S'91-M'00) received the B.Sc. degree in electrical and electronics engineering, and the B.Sc. degree in physics, from Boğaziçi University, Istanbul, Turkey; and the M.S. and Ph.D. degrees in electrical and computer engineering from Wireless Information Network Laboratory (WINLAB), Rutgers University, New Brunswick, NJ. Commencing fall 2010, for three semesters, she was a P.C. Rossin Assistant Professor at the Electrical Engineering and Computer Science Department, Lehigh University, PA. In 2002, she joined the faculty of The Pennsylvania State University, University Park, PA, where she was an Assistant Professor, then Associate Professor, and is currently Professor of Electrical Engineering since 2010. During the academic year 2008-2009, she was a Visiting Associate Professor with the Department of Electrical Engineering, Stanford University, CA. Her research interests are in information theory, communication theory and network science, with recent emphasis on green communications and information security. She received the NSF CAREER award in 2003.

Dr. Yener served as the student committee chair for the IEEE Information Theory Society 2007-2011, and was the co-founder of the Annual School of Information Theory in North America co-organizing the school in 2008, 2009 and 2010. She currently serves on the board of governors as the treasurer of the IEEE Information Theory Society.